\documentclass{aastex}
\usepackage{float}
\usepackage{hyperref}
\usepackage{natbib}
\usepackage{amsmath,amsfonts,amssymb,amscd,amsthm,xspace}
\begin{document}

\title{Magnitude and timing of the giant planet instability: A reassessment from the perspective of the asteroid belt } 

\author{\textbf{A. Toliou$^{(1),(2)}$, A. Morbidelli$^{(2)}$, K. Tsiganis$^{(1)}$}\\  
(1)  Department of Physics, Aristotle University of Thessaloniki, GR-54124 Thessaloniki, Greece. \\
(2) Laboratoire Lagrange, UMR7293, Universit\'e de Nice Sophia-Antipolis,
  CNRS, Observatoire de la C\^ote d'Azur. \\ 
}

\begin{abstract}
It is generally accepted today that our solar system has undergone a phase during which the orbits of the giant planets became very unstable. In recent years, many studies have identified traces of this event and have provided reasonable justification for this occurrence. The magnitude (in terms of orbital variation) and the timing of the instability though (early or late with respect to the dispersal of the gas disk) still remains an open debate. The terrestrial planets seem to set a strict constraint: either the giant planet instability happened early, while the terrestrial planets were still forming, or the orbits of Jupiter and Saturn had to separate from each other impulsively, with a large enough `jump' in semimajor axis \citep{Brasser2009,KaibChambers2016} for the terrestrial planets to remain stable. Because a large orbital jump is a low probability event, the early instability hypothesis seems to be favored. However, the asteroid belt would also evolve in a different way, assuming different instability amplitudes. These two constraints need to match each other  in order to favor one scenario over the other. Considering an initially dynamically cold disk of asteroids, \citet{Morby2010} concluded that a comparably large jump is needed to reconstruct the current asteroid belt. Here we confirm the same conclusion, but considering an asteroid population already strongly excited in eccentricity, such as that produced in the Grand Tack scenario \citep{GrandTack2011}. Because the asteroids existed since the time of removal of the gas disk, unlike the terrestrial planets, this constraint on the width of the giant planet jump is valid regardless of whether the instability happened early or late. Hence, at this stage, assuming an early instability does not appear to provide any advantage in terms of the probabilistic reconstruction of the solar system structure.

\end{abstract}

\section{Introduction}

The orbital structure of the outer solar system can be reproduced in numerical simulations in which the giant planets undergo a phase of orbital instability. This process explains the current orbits of the giant planets \citep{Tsiganis2005,NesvornyMorby2012}, the existence and the properties of the irregular satellites of these planets \citep{Nesvorny2007}, of the Trojans of Jupiter \citep{Morby2005,Nesvorny2013} and of Neptune \citep{NesvornyVokrouhlicky2009}, of the Kuiper belt \citep{Levison2008,Nesvorny2015a,Nesvorny2015b}, of the scattered disk and the Oort cloud \citep{BrasserMorby2013}. 

This list of successes provides compelling evidence for the giant planet instability, however, when the instability actually happened is still a subject of debate. A priori the instability could have happened early (i.e., soon after the dissipation of gas from the protoplanetary disk) or late, depending on the distribution of planetesimals around the giant planet orbits \citep{Levison2011}. By triggering an impact shower onto the terrestrial planets and the Moon, a late instability would have the advantage of explaining the origin of the latest lunar basins, such as Imbrium and Orientale, dating 600-700My after lunar formation \citep{Gomes2005,Bottke2012,Morby2012}. In fact, the formation of these basins is very unlikely to be due to planetesimals leftover from the terrestrial planet formation process \citep{Bottke2007}. 

Nevertheless, there are other complications. If the giant planet instability happened late, the terrestrial planets were already formed. Thus an additional constraint is that the terrestrial planets must have avoided being destabilized or overexcited. \citet{Brasser2009} showed that this is possible if the orbits of Jupiter and Saturn separated from each other in an impulsive manner as a result of encounters with another planet. They dubbed this kind of evolution ``jumping-Jupiter''. The period ratio between Saturn and Jupiter has to jump fast across the 2.1-2.3 range, so that the $g_1=g_5$ resonance is relocated inside of the orbit of Mercury without sweeping through the terrestrial planet region. This ensures that the eccentricities of the terrestrial planets will not have enough time to grow by resonating with the giant planets.

In a recent paper, \citet{KaibChambers2016} performed 200 simulations of the giant planet instability phase and they found that only one simulation successfully recovered, within a reasonable approximation, the current orbits of the giant planets while avoiding the excitement of the angular momentum deficit (AMD; see \citet{Laskar1997,Chambers2001} for a definition) of the terrestrial planets beyond its present value; that is, even an impulsive evolution for Jupiter and Saturn has a very small probability of giving the correct evolution. Thus, they concluded, the giant planet instability is more likely to have occurred early, while the terrestrial planets were still growing. Then, even in the case of a small `jump' that would induce resonant excitation, the growing planets would recover, as the orbits of the embryos would be continuously damped by dynamical friction  exerted by the planetesimals \citep{Chambers2001,O'Brien2006}. 

The success rate of the simulations of  \citet{KaibChambers2016} may be particularly low because the terrestrial planets were initially assumed to be on circular and coplanar orbits. In fact, in this case, the AMD can only grow during the giant planet instability, whereas the AMD can also decrease, for some combinations of secular phases, if the terrestrial planets have initially none-zero eccentricities and inclinations \citep{Brasser2009}. \citet{NesvornyMorby2012} computed a probability of 7-10\% for Jupiter and Saturn to have a jump of appropriate amplitude,  in simulations where they initially assumed three planets of mass comparable to Uranus and Neptune, where one of these was eventually ejected. Using one of these successful simulations, \citet{Brasser2013} concluded that, if the AMD of the terrestrial planets is initially 70\% of the current value, the probability that the AMD remains below the current value is $\sim20$\%. Combined together, these works suggest that the overall probability of success is 1--2\%. Consistently, \citet{RoigNesvorny2015} presented several simulations that are successful in reproducing both the orbits of the giant and terrestrial planets. Nevertheless, a success probability of a few percent, although higher than in \citet{KaibChambers2016}, remains puzzlingly small and supports their argument in favor of an early instability of the giant planets, where a more probable `small' jump, followed by a phase of slow migration of the giant planets, would likely not affect the stability of the forming terrestrial planets.

The terrestrial planets, however, are only part of the story. The asteroid belt is the other (fundamental) part. If the jump in period ratio between Saturn and Jupiter is too short, the $g=g_6$ resonance, which affects the eccentricities of the asteroids, lands in the asteroid belt, and then, while the giant planets migrate smoothly toward their current orbits, this resonance sweeps through the inner belt strongly affecting bodies at low inclinations. This severely depletes the asteroids at low inclinations in the inner belt but not those at high inclinations, which are not swept by the resonance. As a consequence, the final ratio between asteroids above and below the $g=g_6$ resonance line results to be much larger than the observed ratio \citep{Morby2010}. Starting with an asteroid belt without high inclination asteroids does not solve the problem because the $s=s_6$ resonance, which affects the inclinations of the asteroids, sweeps through the belt before the $g=g_6$ resonance, so that many asteroids are kicked to high inclination before the $g=g_6$ resonance passes through the inner belt \citep{Morby2010}.

As mentioned above, previous studies by \citet{Brasser2009}, \citet{Brasser2013}, and also \citet{AgnorLin2012} have shown that damping the AMD of the planets is much less probable than increasing it. However, there is an important difference when discussing the asteroids' population. The main belt contains a population of asteroids of which the number does not remain constant over hundreds of Myr, as a fixed set of planets does. Consequently, we cannot simply consider an initial average AMD of the asteroid belt and compare it with the final AMD, since the population will not be the same. True, the average AMD will most likely increase (as indicated by the aforementioned works), but the real question is whether, at the end, there will be more or fewer low-eccentricity asteroids than at the beginning. Essentially, if the initial population of high-eccentricity asteroids is large, the total number of asteroids implanted at low eccentricities, even if it is a low probability event, may be significant. From this discussion it is clear that the constraints set by both populations, that is planets and asteroids, have to be considered simultaneously

The jump in period ratio between Saturn and Jupiter has to be as large as required by the constraint provided by the AMD of the terrestrial planets system, if we assume the terrestrial planets are already formed. This constraint is no longer strict, if we accept the early instability hypothesis, however, the same argument would not apply to asteroids. In fact, all asteroids were fully formed by the time the gas was dispersed from the protoplanetary disk and there have been no known inclination damping processes that occurred since then. In other words, the timing of the instability (early or late) is not the key parameter for asteroids. It is the magnitude of the instability that can be critical. Any constraint on that would have to hold, no matter when the instability happened.

In principle, one may imagine that, if the sweeping of secular resonances through the inner belt was early enough, the still ongoing process of terrestrial planet formation could reshuffle the inclination distribution in the inner belt, but \citet{WalshMorby2011} explicitly investigated and excluded this possibility. \citet{WalshMorby2011}, however, considered a disk of asteroids initially on low eccentricity and low inclination orbits. Since then, a new scenario of the evolution of the inner solar system has been developed: the Grand Tack scenario \citep{GrandTack2011}.  In this scenario, the migration of Jupiter in the disk of gas drove the planet across the asteroid belt leaving the asteroids, at the end of the gas-dissipation phase, on orbits very excited in eccentricity. An asteroid eccentricity distribution skewed toward high values opens a new possibility: the number of high-eccentricity asteroids whose eccentricity was decreased by the sweeping of the $g=g_6$ resonance might be high enough so that, at the end of the process, the number of low inclination asteroids is substantial enough and no apparent deficiency at low inclinations is observed in the inner belt. If this were true, this would  relax the constraint set by the inner asteroid belt on the amplitude of the jump of the Saturn-Jupiter period ratio, dramatically increasing  the likelihood of successful evolutions in the early-instability framework, thereby avoiding the severe constraint set by the AMD of the terrestrial planets in the late instability framework \citep{KaibChambers2016}. 

Given its potential importance, in this paper we study this process thoroughly. In Section~2 we describe the methods we used and the initial conditions we adopted. In Section~3 we present our first results, starting from the final asteroid distribution provided by the Grand Tack simulations of \citet{GrandTack2011}. Because a recent work \citep{Deienno2016} argues that the inclination distribution at the end of these Grand Tack simulations was too excited, in Section~4 we consider only the asteroids whose initial inclination is below 20 degrees. The conclusions of this work are discussed in Section~5.

\section{Numerical methods and initial conditions}
In this study, we investigate the orbital evolution of asteroids in the main belt under the hypothesis that Jupiter and Saturn had a short orbital jump, followed by a long-ranged planetesimal-driven migration. Besides Jupiter and Saturn, our model includes also prototerrestrial planets, planetary embryos, and a populous disk that extends between $0.5<a<3.5$~AU in semimajor axis. 

\subsubsection{Asteroids}
The orbital elements of the asteroids were chosen to be in accordance with the end state of the Grand Tack model \citep{GrandTack2011}. Specifically, we considered 40 simulations from \citet{JacobsonMorby2014} that gave the best results in terms of the final terrestrial planets and timing of the Moon forming impact. We selected all planetesimals that have orbits with semimajor axis $1.6<a<3.5$~AU at the end of the inward and outward migration of Jupiter and had aphelia inside Jupiter's orbit.

We have constructed ten sets of initial conditions of the post-Grand Tack asteroid belt. Each set consists of the same 4600 particles that satisfy the conditions described above, but differ for the orbits of the protoplanets and embryos in the inner disk (see below). All the particles in the $1.6<a<3.5$ AU range have a mass of $m_{ast}=3.8 \cdot 10^{-6}\,M_{\oplus}$, so that the total mass in that region corresponds to the average mass that we find in the Grand Tack simulations.

\subsubsection{Embryos and protoplanets} 
Our ten sets of initial conditions differ in the distribution of embryos and planetesimals inside 1.6~AU, which were also taken from \citet{JacobsonMorby2014}. Among their 40 best simulations, we considered those that assume a ratio of 8 between the total masses in embryos and planetesimals and an individual embryo mass of 0.8 Mars masses. We adopted as initial conditions the distribution of objects at the end of the outward gas-driven migration of Jupiter. This arrangement leads to a total of $\sim5500$ bodies per set.

\subsubsection{Giant planets}
The simulations were performed with the N-body integrator SyMBA \citep{Symba1998} that treats close encounters between massive bodies in a symplectic manner. In our case, the giants planets, terrestrial protoplanets, and embryos can feel each other's gravity and can also interact gravitationally with the asteroids. We do not account for self-gravity between asteroids. In addition, in case of a collision, we allow merging between all bodies except between asteroids.

In this work we assume that Jupiter and Saturn had a short orbital jump, followed by extensive planetesimal-driven migration. At the end of the Grand Tack model, Jupiter and Saturn were in their mutual 2:3 mean motion resonance and had quasi-circular and quasi-coplanar orbits. We enact a short jump by setting the initial orbits of these planets at $a=5.4$~AU, $e\simeq0.04$, $i\simeq1.71^{\circ}$ and $a=8.7$~AU, $e\simeq0.07$, $i\simeq1.03^{\circ}$ respectively. In this way the planets are placed beyond their mutual 1:2 resonance. For the planetesimal-driven migration phase we use the \citet{Malhotra1995} recipe, in which the semimajor axis changes exponentially as 
\begin{equation}
a(t)=a_f-\Delta a \; exp(-t/\tau)                                                                                                                                                                                                   ,\end{equation} 
   
where $\Delta a$ is the difference between the initial and final a value. In order to achieve this, we implement an additional acceleration in the SyMBA integrator along the direction of velocity $\hat{\rm{v}}$, that is equal to 
  
\begin{equation}
\Delta\mathbf{\ddot{r}}=\frac{\hat{\rm{\textbf{v}}}}{\tau} \left( \sqrt{\frac{GM_\odot}{a_f}} - \sqrt{\frac{GM_\odot}{a_i}} \right) exp\left( - \frac{t}{\tau} \right)                                                                                                                                                                                                 ,\end{equation} 
  
where $a_i$ is the initial semimajor axis and $a_f$ is the final semimajor axis we want to reach. This force acts upon both giant planets. No additional forces damping the orbits of the planets were used.

\begin{figure}[H]
 \includegraphics[width=\hsize]{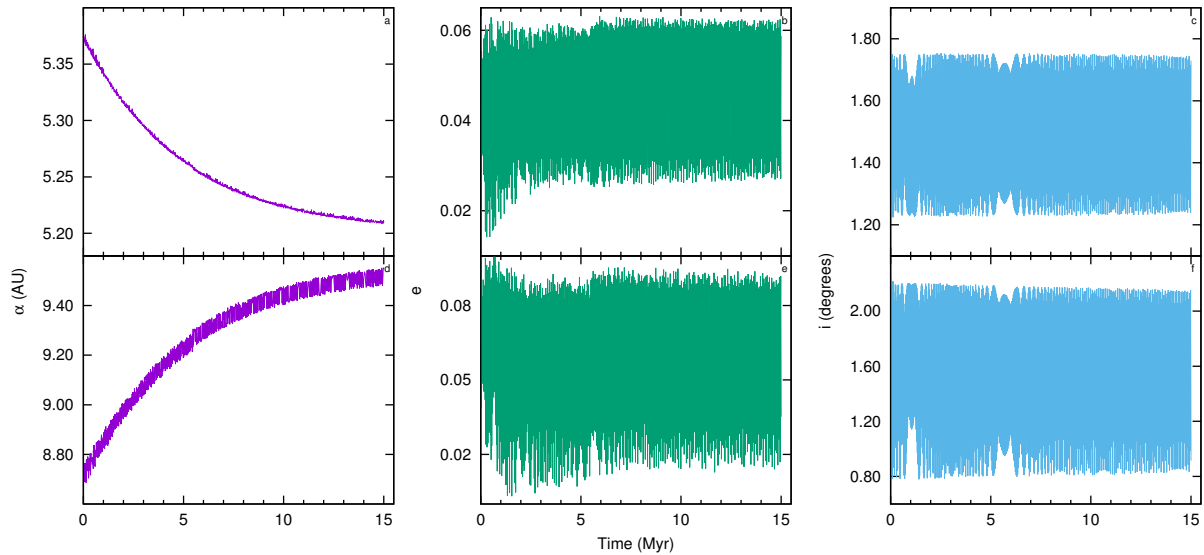}
 \caption{Example of Jupiter and Saturn's typical migrational evolution: Jupiter's  evolution in time of :(a) semimajor axis, (b) eccentricity, (c) inclination and Saturn's evolution in time of (d) semimajor axis, (e) eccentricity,  and (f) inclination.}
 \label{fig:planets}
\end{figure}

We adopt $\tau=5$~Myr as an appropriate value of timescale for migration driven by planetesimal scattering, as described in \citet{Morby2010}. All ten runs use the same initial conditions and migration parameters for the planets. The typical orbital evolution (see \autoref{fig:planets}) suggests that the migration process is smooth, the planets do not cross any strong MMRs (e.g., the 2:5), and they have final values of semimajor axis, eccentricity, and inclination that are very close to the current values.

\section{Results}

We present the results of ten simulations, all of which have a total integration time of $t_{tot}=15$~Myr. This timespan is equal to three times the e-folding time, which translates to $95\%$ of the migrating distance being covered. This is enough for secular resonances to sweep through the post-Grand Tack main belt. The goal of this study is to investigate the effect that this particular mechanism has on the structure of the belt. 

\begin{figure}[H]
\centering
 \includegraphics[scale=0.65]{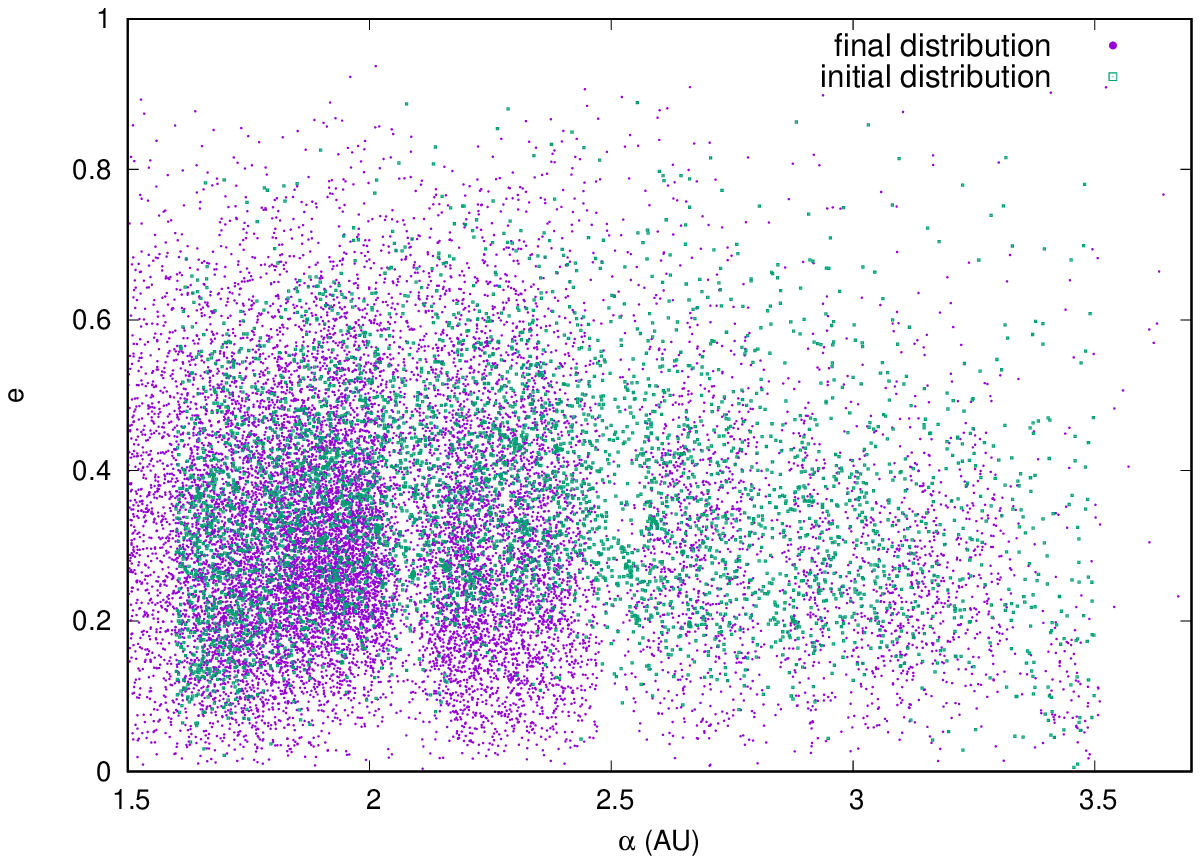}\includegraphics[scale=0.65]{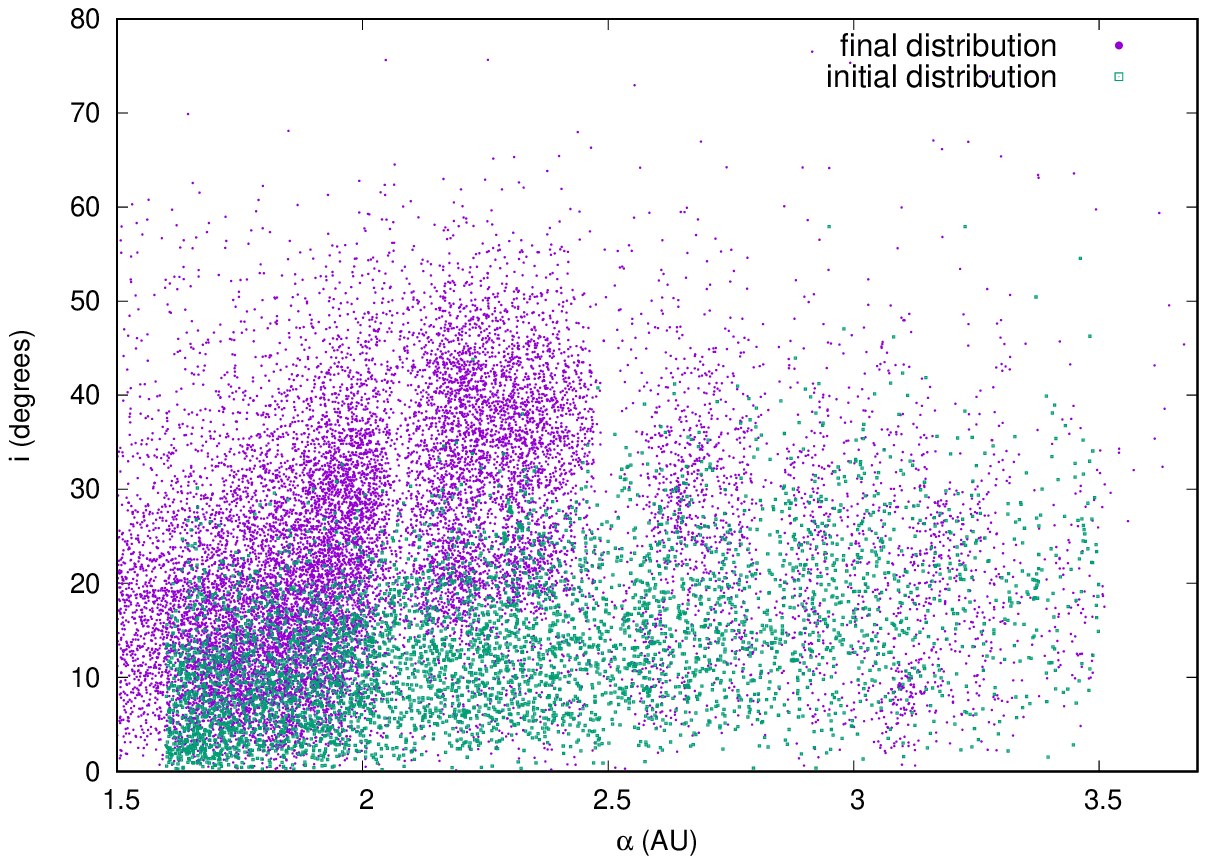} \\
 (a) \hspace{7.7cm} (b)
 \caption{(a) Initial and final a,e distribution and (b) initial and final a,i distribution of all asteroids for all 10 runs combined}
 \label{fig:ae-asteroids}
\end{figure}

In \autoref{fig:ae-asteroids} we can see the initial and final distribution on the semimajor axis - eccentricity plane, as well as on the semimajor axis - inclination plane of all asteroids for all runs combined. In the following we only consider the particles with $q>1.8$~AU because those particles would remain in the asteroid belt, while the others would ultimately be removed by encounters with the terrestrial planets.

\begin{figure}[H]
\centering
  \includegraphics{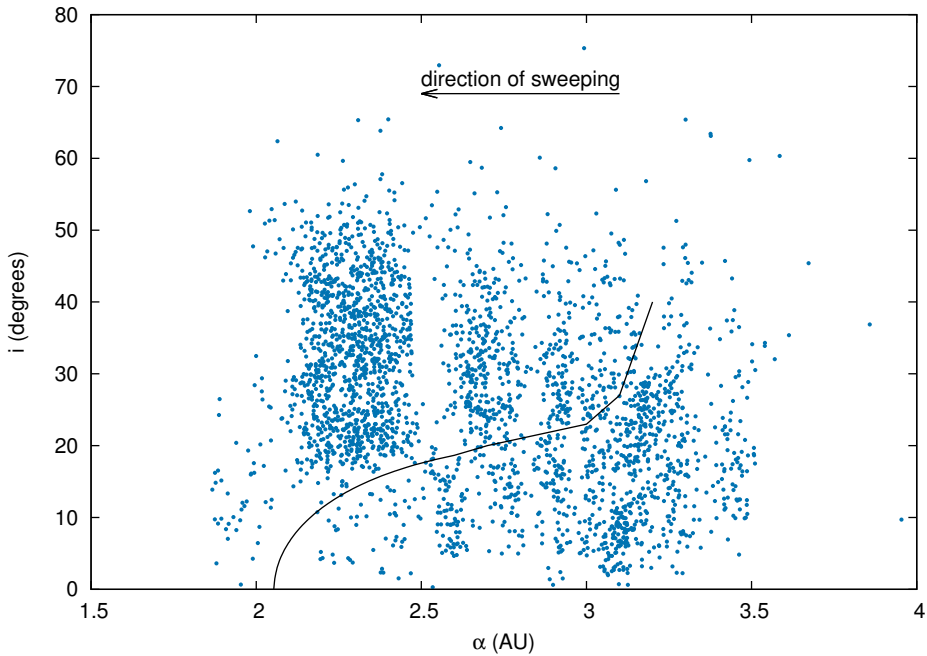}
  \caption{Final a,i distribution of all asteroids which belong to the main belt, i.e. with $q>1.8$~AU. The black line shows the location of $g=g_6$.}
  \label{fig:ai-mainbelt}
\end{figure}

When we focus on the semimajor axis - inclination plane, we see that the distribution of these particles is incompatible with the current structure. Specifically, we notice that the inner part of the asteroid belt is strongly depleted of low inclination bodies (see \autoref{fig:ai-mainbelt}).

The mechanism that leads to such an excited belt has been identified to be the sweeping of the $g=g_6$ secular resonance \citep{Morby2010}. $g_6$ is the eigenfrequency of the solar system that is associated with Saturn's rate of precession of longitude of pericenter $\dot{\varpi}$ and its location depends on Jupiter and Saturn's semimajor axis. Consequently, when the planets migrate, the location of the resonance also migrates, sweeping through the main belt toward $a\sim2$~AU and primarily affecting low inclination asteroids. Although the resonance can either increase or decrease the eccentricities of the already excited post-Grand Tack main belt, depending on the longitude of perihelion of each asteroid at that time of resonance sweeping, the outcome suggests that ejection to high eccentricities was more favorable.

We selected from the AstDys catalogue all asteroids with perihelion distance $q>1.8$~AU, semimajor axis $a<2.5$~AU, and diameter approximately larger than 50~km, to compare the structure of the inner main belt obtained from our simulations to the actual observed asteroids that reside in that region in more detail. 

We only selected objects of this size because they are not strongly affected by family formation events and they are large enough to avoid having their orbits altered by Yarkovsky drift during the following 4 Gyr of evolution. In addition, they are not subject to obervational bias, as all objects with these absolute magnitudes have likely been discovered \citep{Jedicke2002}. As a result, they serve as a valid source of information for the structure of the main belt at the era we study.  

We considered two cases for the upper limit of the absolute magnitude $H$ because of the variation of albedos for asteroids of different spectral types. Although $H<9$ could be sufficient given the fact that at $a<2.5$~AU most asteroids are S-type, we also regarded $H<10$ to account for C-type objects as well.

\begin{figure}[H]
\centering
  \includegraphics{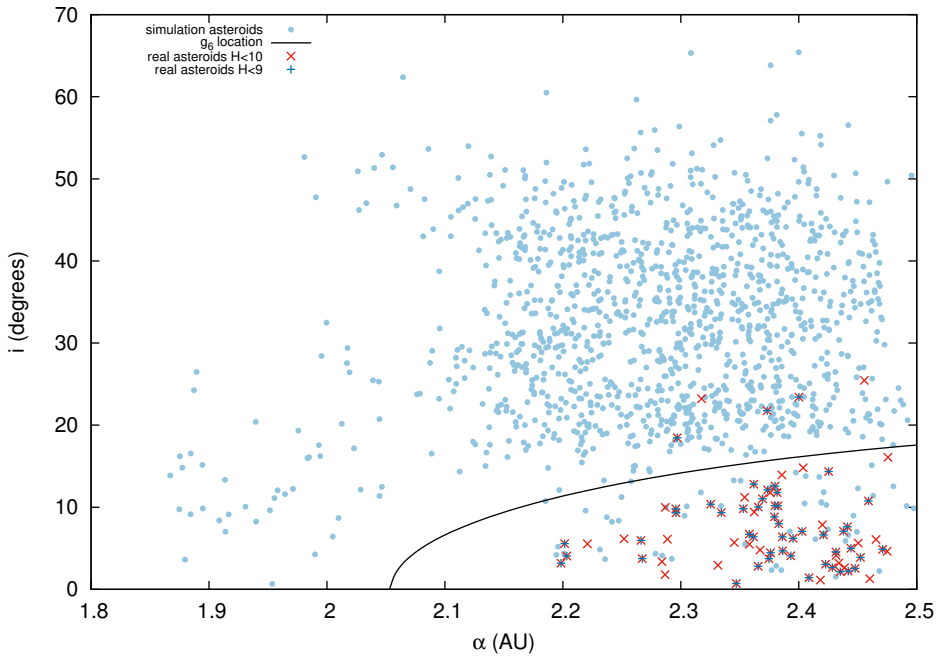}
  \caption{Final a,i distribution of main belt asteroids with semimajor axis $a<2.5$~AU. The whole population of asteroids of all ten runs is shown in light blue. Real asteroids with $H<9$ and $H<10$ are colored by dark blue and red points, respectively. The black line illustrates the current location of $g=g_6$}
  \label{fig:nu6_all}
\end{figure}

In \autoref{fig:nu6_all} we focus on the inner belt. We illustrate the inclination distribution of all our particles from all ten runs combined, after evolving the system for 15 Myr, when the giant planets have practically stopped migrating. We compare this distribution with the inclination distribution of real asteroids for both limiting values of $H$.

Apparently, the difference between the simulated and the observed asteroids is vast. The low inclination band of the synthetic population has suffered great depletion, while the real population has very few asteroids at higher inclinations, in fact, above the $g=g_6$ line.   

An indicator that is very easy to obtain for the inclination distribution of inner main belt asteroids is to count the number of asteroids above or below $g=g_6$  and calculate their ratios, as applied in \citet{Morby2010}. Out of a total of 1276 particles with $a<2.5$~AU and $q>1.8$~AU, only 51 end up on orbits below the $g=g_6$ line, giving a percentage of $4\%$. On the other hand, real asteroids in the same region are $93.88\%$ and $93.42\%$ of the total population in the same range of $a$ for $H<9$ and $H<10,$ respectively. In \autoref{tab:ratios} we present the detailed percentages for all ten runs.

\begin{table}[H]
\centering
\caption{Percentage of simulated with $a<2.5$~AU AND $q>1.8$~AU that are \textbf{below} the $g=g_6$ secular resonance line for each run seperately, for particles of all runs combined, as well as the percentage of the real asteroids with radii $H<9$ and $H<10$. The third column lists the percentages for asteroids with initial inclinations $i\leq20^{\circ}$ and the fourth column lists those modified after 200~Myr of evolution. }
\label{tab:ratios}
\begin{tabular}{c c c c }
\hline\hline
Simulation & \% below $g=g_6$ & \begin{tabular}[c]{@{}c@{}} \% below $g=g_6$ \\ $(i\leq20^{\circ})$ \end{tabular} & \begin{tabular}[c]{@{}c@{}} \% below $g=g_6$ \\ (t=200 Myr) \end{tabular} \\ 
\hline     
run1                & 3.54    &   4.55  &   7.69     \\ 
run2                & 3.15    &   2.83  &   0.00     \\ 
run3                & 2.31    &   1.89  &   0.00     \\ 
run4                & 7.64    &   8.62  &  23.81     \\ 
run5                & 2.94    &   1.25  &   0.00     \\ 
run6                & 4.35    &   4.63  &  15.00     \\ 
run7                & 1.68    &   1.06  &   0.00     \\ 
run8                & 5.08    &   5.15  &   9.09     \\ 
run9                & 6.15    &   6.86  &   0.00     \\ 
run10               & 2.36    &   2.27  &   0.00     \\ 
total               & 4.00    &   4.07  &   8.44     \\ 
real $H<9$          & 93.88   & ... & ...                 \\ 
real $H<10$         & 93.42   & ... & ...                 \\ 
\hline
\end{tabular}
\end{table}

It is clear though that these relative numbers could change during the following $\sim100$~Myr as the system of terrestrial planets is forming. Thus, after integrating for the first 15 Myr, we continued the simulations and extended the total integration time to a few hundred Myr to examine the effects that embryos and prototerrestrial planets, which were left in the terrestrial zone, have on the inner main belt asteroids. We resumed all ten runs until 200 Myr, but only included the giant planets, the remaining (after mergers and ejections) embryos and protoplanets, and the inner belt particles that had survived the first 15 Myr. A total of 1276 particles are contained in the asteroid belt region in all runs combined.

\begin{figure}[H]
\centering
  \includegraphics{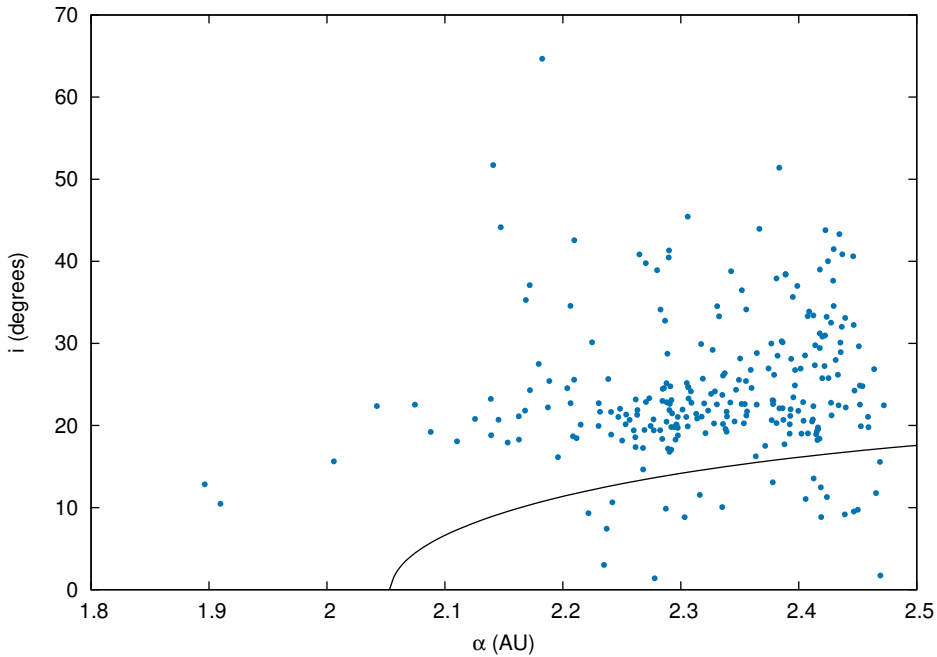}
  \caption{a,i distribution of main belt asteroids with semimajor axis $a<2.5$~AU after integrating for $t=200$~Myr. The black line illustrates the location of $g=g_6$.}
  \label{fig:final}
\end{figure}

In \autoref{fig:final} we see the final distribution of inner belt asteroids that survive. The long-term effects have not imposed a qualitatively different structure, except for lowering the maximum inclination limit to $\sim40^{\circ}$. These results are in agreement with those of \citet{WalshMorby2011}. The continuation of the simulation led to 237 remaining objects, of which only 20 are below the $g=g_6$ line, resulting in a new percentage of $8.44\%$.

\begin{figure}[H]
\centering
  \includegraphics[scale=0.65]{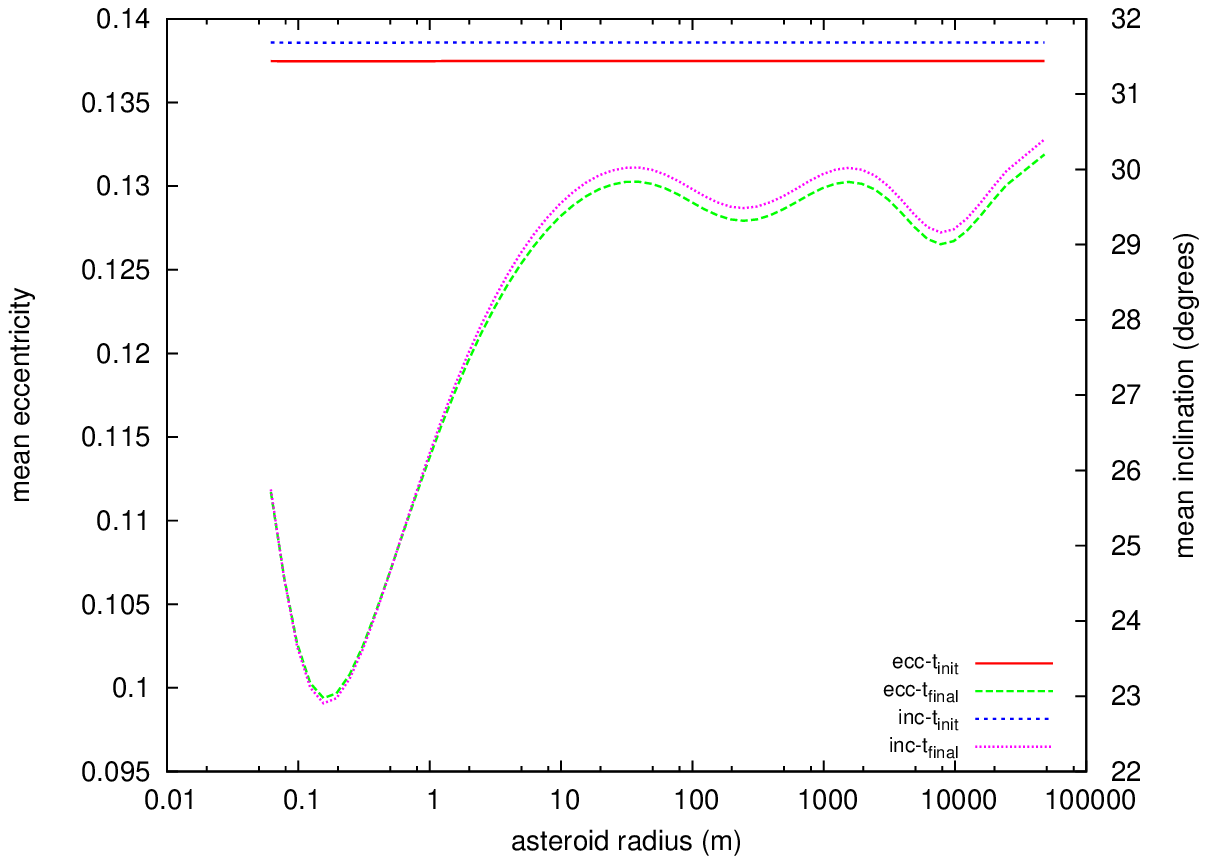}\includegraphics[scale=0.65]{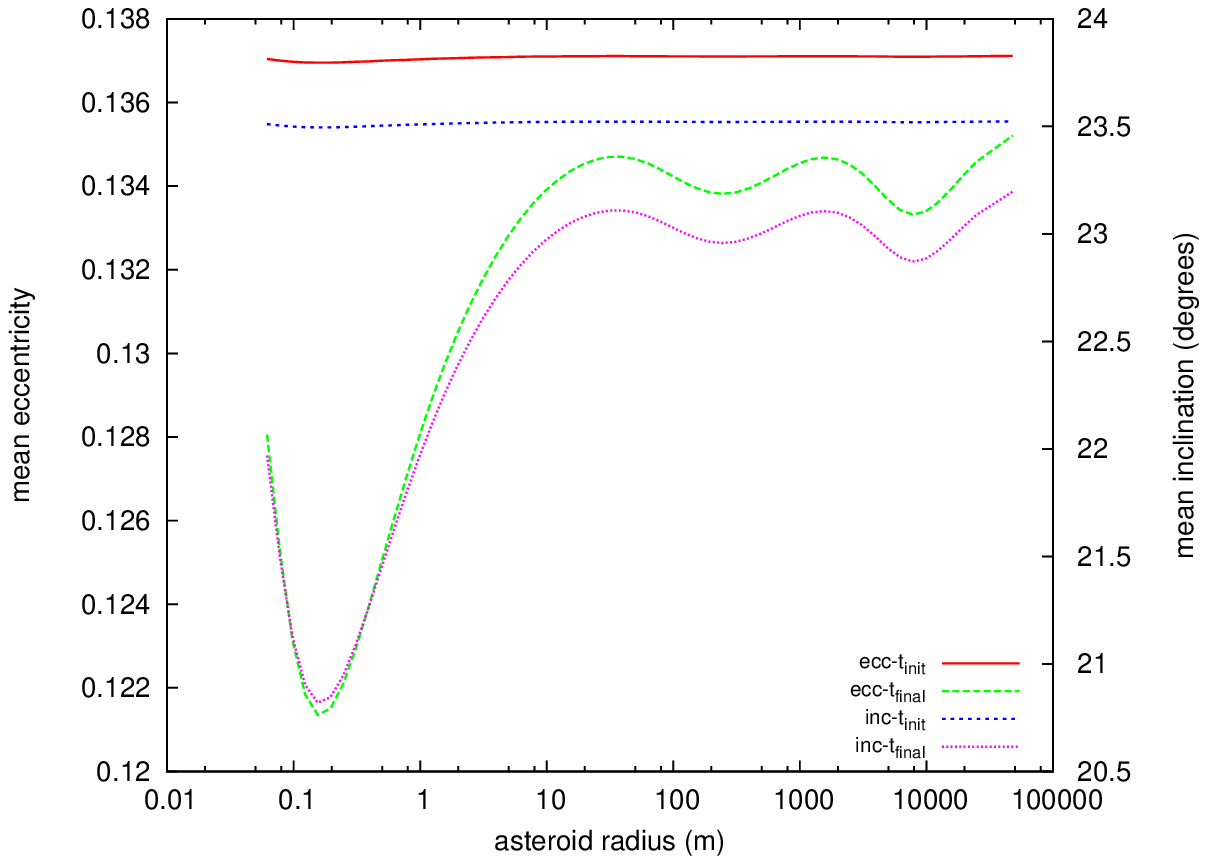} \\ 
  (a) \hspace{7.7cm} (b)
 \caption{(a) Here is the mean value of eccentricity and inclination after simulating a collisional evolution of $\sim4$~Gyr with respect to asteroid size starting at an (a) $t_{initial}=15$~Myr and (b) at $t_{initial}=200$~Myr.}
 \label{fig:collisions}
\end{figure}

Another mechanism that could possibly change the final orbital distribution of our simulated main belt asteroids is the effect of collisions between them. Hence, we investigated the possibility of removing high inclination asteroids through collisions during 4~Gyr, using a short version of the \rm{BOULDER} code \citep{Morby2009}. The code was modified by suppressing the evolution of the size frequency distribution of the asteroids and also, by considering only the collisional damping term in the equations describing the evolution of eccentricities and inclinations. Essentially, we studied the collisional evolution of the inner belt in both the time span between 15~Myr - 4~Gyr as well as between 200~Myr - 4~Gyr, that is after our assumed $\sim185$~Myr phase of planet formation. In \autoref{fig:collisions} we can see that for medium-sized asteroids and larger (i.e., all objects with radii $R>100$~m), the mean values of eccentricity and inclination are not significantly altered.

More specifically, in the first case we used, as initial conditions, a population with mean values identical to those of the remaining 1276 objects at the end of the 15~Myr simulations: $\bar{e}=0.14$ and $\bar{i}=31.7^{\circ}$. The total mass that this population represented was safely estimated to be $\sim10$ times the current asteroid belt mass by taking into account the planet formation process in the terrestrial region (that removes a lot of mass), as well as the subsequent effect of chaotic diffusion that would decrease the mass of the belt by $\sim2$ throughout 4~Gyr of evolution. We observe that collisional damping cannot reduce the number of high inclination asteroids, as medium and larger size asteroids have on average an inclination of $\bar{i}\simeq25^{\circ}$, which is nowhere close to the current value of $\simeq10^{\circ}$.  In addition, assuming that the inner part only accommodates one-third of the total mass of the main belt, should we consider the mass distributed homogenously, we can choose the total mass of the population to be $\sim3$ times the current asteroid belt mass. We then observe that the average inclination is $\bar{i}\simeq29.8^{\circ}$, which is much farther from the observed value.

The same holds true for the second experiment, in which our initial population has $\bar{e}=0.14$ and $\bar{i}=23.5^{\circ}$ and is assumed to start evolving after $\sim185$~Myr, when planet formation has ceased. The total mass selected was $\sim4$ times  the current main belt mass to account for the depletion during planet formation (still a safe value). The new mean inclination at $t=4$~Gyr is $\bar{i}\simeq21.5^{\circ}$ and in the case in which we also consider only one-third of the total mass, this value becomes $\bar{i}\simeq 23^{\circ}$; these values are both still very far from the observed.

\section{Low inclination asteroids}

We also investigated the final distribution of the particles that had initially (i.e., at the end of the Grand Tack phase) inclinations $i\leq20^{\circ}$. The reasoning behind this choice comes from the fact that the Grand Tack scenario seems to give an inclination distribution that is somewhat too excited. In fact, \citet{Deienno2016} showed that, even in the case of a big jump of Jupiter (so that $g=g_6$ sweeping through the asteroid belt is avoided), there are still too many asteroids above the $g=g_6$ resonance line. This problems goes away if one assumes that the Grand Tack inclination distribution was truncated at 20 degrees \citep{Deienno2016}. 

\begin{figure}[H]
\centering
\includegraphics{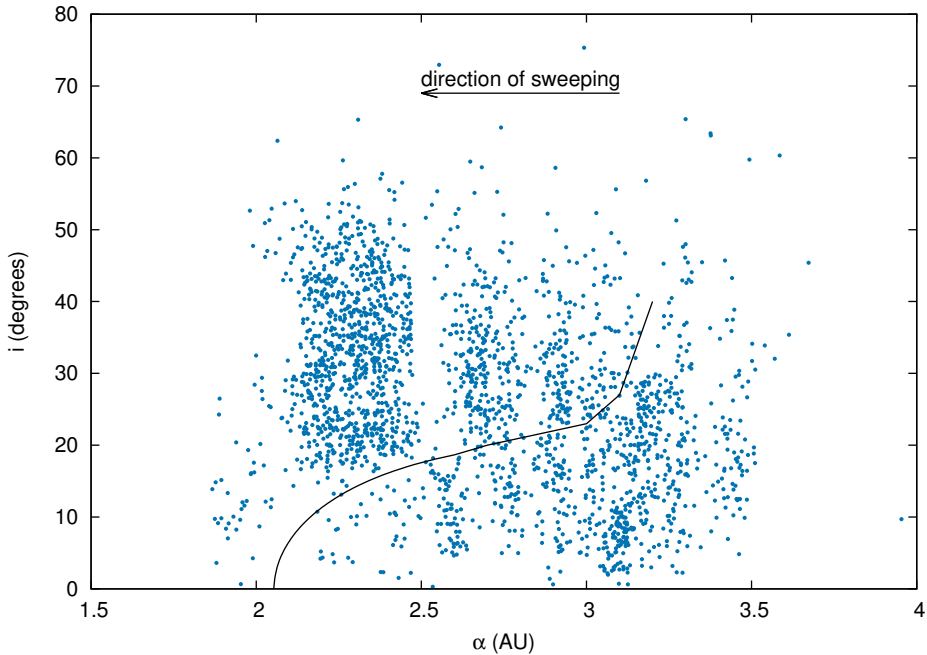}
\caption{Final a,i distribution (at $t=15$~Myr) of all simulated asteroids that belong to the main belt and had initial inclination $i\leq20^{\circ}$. The black line shows the current location of $g=g_6$}
\label{fig:ai_mblt20}
\end{figure}

In \autoref{fig:ai_mblt20} we present the final distribution of these particles on the semimajor axis - inclination plane. We see that even assuming an initial $i\leq20^{\circ}$ does not lead to a more populous inner belt in the low inclination range. 
This is different from the results of \citet{Deienno2016}, precisely because of the smooth migration of Jupiter and Saturn that we impose in this case.
The blame is to be cast upon the $s=s_6$ secular resonance.  This resonance corresponds to the rate of the regression of the longitudes of Jupiter and Saturn, $\dot{\Omega}$, and plays an important role in sculpting the main belt, by affecting the inclinations of the asteroids. \citet{Gomes1997} finds that in the migration scenario that we are studying $s=s_6$ sweeps the belt from 2.8~AU until it reaches its current location at 1.9~AU.

More specifically, at the inner belt region where we focus on, $s=s_6$ sweeps first, raising inclinations before $g=g_6$ sweeps through. As a consequence, many asteroids acquire a much higher inclination than their initial $i\leq20^{\circ}$ and when $g=g_6$ reaches their semimajor axis, their eccentricities remain unaffected. However, low inclination asteroids are not very lucky as they suffer the same fate we described previously; their eccentricities are raised, which forces them to cross the terrestrial planet region and either merge with the embryos or get ejected. The percentage of the total population below the $g=g_6$ line becomes $4.07\%$ (out of 1007 asteroids only 41 are below the resonance line). In \autoref{tab:ratios} we also show the new percentages derived from these simulations.

Hence, irrespective of whether initially asteroids had $i\leq20^{\circ}$ or not, the slow sweeping of secular resonances would result in an inner-belt population that is very different from the observed one (i.e., very few low inclination bodies at $i\leq20^{\circ}$). Of course, assuming collisions as in the previous case, could not remedy this result.

\section{Discussion and conclusions}

In this paper we investigate how the asteroid belt was sculpted by the evolution of the giant planets that followed their phase of dynamical instability. Previous works \citep{Morby2010,WalshMorby2011} did a similar investigation considering an asteroid disk initially on dynamically cold orbits. Here we consider instead an asteroid belt that is initially very excited in eccentricity (and inclination), such as the one resulting from the Grand Tack scenario. The scope of this study is to shed light on the problem of disentangling the correct instability `magnitude' and `timing' that could result both into a set of stable terrestrial planets and an asteroid belt that looks like the current belt.

During the phase of dynamical instability the giant planets suffered close encounters with each other. This likely led to an impulsive separation of the orbits of Jupiter and Saturn due to encounters with another planet (Uranus, Neptune or a putative third ice-giant planet  \citep{NesvornyMorby2012}). This jump in the semimajor axis of Jupiter and Saturn was followed by a phase of smooth planetesimal-driven migration, further increasing the orbital separation between the two major planets. As explained in the introduction, the question really is what was the amplitude of the jump relative to the range of smooth, planetesimal-driven migration. 

We have found that if the jump had been short, bringing Jupiter and Saturn just beyond their mutual 1:2 mean motion resonance, the smooth planetesimal-driven migration that should have followed this jump, in order to bring the giant planets to their current orbital separation, would have had devastating effects on the inner asteroid belt. This is because, in this case, two secular resonances, $s=s_6$ and $g=g_6$ would have swept through the inner asteroid belt. The first resonance would have raised the inclinations of many asteroids and the second would have removed, by exciting their eccentricities, most of the asteroids that remained at low inclination. Consequently, the final ratio between the number of asteroids with inclination above and below the  $g=g_6$ resonance would have been much larger than the observed ratio. This result holds regardless of the initial inclination distribution of the asteroids. In this sense, it becomes evident that to reproduce our results, it is not necessary to take an orbital distribution that is identical to the post-Grand Tack distribution as an initial condition for the asteroids.  The same conclusions hold true simply by considering a sufficiently excited main belt.

\citet{KaibChambers2016} argued that the `jump' in the giant planets' orbits and the associated impulse in eccentricity given to the terrestrial planets was likely not large. Consequently, to ensure the stability of the terrestrial planets, the giant planet instability should have happened early on, when they had not formed yet. The planetesimal disk could then recover from the excitation induced by the sweeping of resonances. However, this probabilistic argument against a large `jump' (and against a late instability) does not take the constraints set by the asteroid belt into account. No model, including the one presented in this paper, succeeds in reproducing the current asteroid belt structure, if the jump between Saturn's and Jupiter's period ratio is small.

In another paper, \citet{Deienno2016} showed that if the jump in the semimajor axis of Jupiter and Saturn was large enough to bring the $g=g_6$ at $i=0^{\circ}$ inward of $\sim2.2$~AU, the asteroid distribution could have evolved toward one similar to the observed distribution. In particular, if the inclination distribution of the asteroids at the end of the Grand Tack phase did not exceed $\sim20^{\circ}$ the final ratio between the asteroids above and below the  $g=g_6$ resonance is about correct. A similar result was found in \citet{RoigNesvorny2015}.

Thus, the results we report here, together with those of \citet{Deienno2016} and \citet{RoigNesvorny2015}, show that Jupiter and Saturn had to have suffered a large semimajor axis jump. Such a large jump is relatively rare in simulations of the giant planet instability; it occurs with a probability of $\sim7$\% in the simulations of \citet{NesvornyMorby2012}. Nevertheless, it has to have happened, otherwise the current distribution of asteroids in the inner belt would be different. This conclusion holds true whatever the timing of the giant planet instability.

Hence, once given this constraint on the amplitude of the jump of the  orbits of Jupiter and Saturn, the probability that the terrestrial planets, if they already existed (i.e., if the instability occurred later than $\sim100$~Myr), did not exceed their current angular momentum deficit (AMD) is actually several tens of percent \citep{Brasser2013}. Then, the low AMD of the terrestrial planets is no longer a strong constraint in favor of an early instability of the giant planets and the subject of the timing needs to be further investigated.

In principle, one might think that, in the following 4~Gyr of evolution, the high inclination part of the inner belt could be eroded by the terrestrial planets, so that only low inclination asteroids survive, resulting in an orbital distribution similar to the current one. However, \citet{Deienno2016} argue that the inclination distribution of the asteroids does not change much during this time span. Nevertheless, even if the ratio between high inclination and low inclination objects is altered significantly, this would not resolve the problem of having too few low inclination objects at $a<2.5$~AU compared to $a>2.5$~AU.

In summary, we can safely conclude that even if a `short jump' is a higher probability solution for the giant planet instability, it does not reproduce the asteroid belt under any assumption for their initial distribution. Hence, even if an early instability appears to be a safer choice for the terrestrial planets (as `small jump' evolutions would also work), it does not lead to any advantage for the asteroid belt, with respect to the late instability hypothesis. The probability of having both populations evolving in the desired way is controlled by the magnitude of the jump, which had to be large, as also shown in this paper.

\section*{Acknowledgements}
A.Toliou wishes to thank OCA for their kind hospitality during her stay there.

\bibliography{referencesv2}

\end{document}